\documentclass{rspublic}

\usepackage{graphicx}

\newcommand{\rmi}{\mathrm{i}}

\begin{document}
\title[On the dragging of light by a rotating medium]{On the dragging of light by a rotating medium}
\author[J. B. G\"otte, S. M. Barnett and M. Padgett]{J\"org B. G\"otte$^1$, Stephen M. Barnett$^1$ and Miles Padgett$^2$}
\affiliation{$^1$Department of Physics, University of Strathclyde, SUPA, John Anderson Building, Glasgow G4 0NG, UK \\
$^2$Department of Physics and Astronomy, University of Glasgow, SUPA, Kelvin Building, Glasgow G12 8QQ, UK}
\label{firstpage}
\maketitle
\begin{abstract}{image rotation, polarisation, rotating dielectric, specific rotary power}
When light is passing through a rotating medium the optical polarisation is rotated. Recently it has  been reasoned that this rotation applies also to the transmitted image (Padgett \textit{et al.} 2006). We examine these two phenomena by extending an analysis of Player (1976) to general electromagnetic fields. We find that in this more general case the wave equation inside the rotating medium has to be amended by a term which is  connected to the orbital angular momentum of the light. We show that optical spin and orbital angular momentum account respectively for the rotation of the polarisation and the rotation of the transmitted image.
\end{abstract}

%
% INTRODUCTION
%

\section{Introduction}
Jones (1976) studied the propagation of light in a moving dielectric and showed by experiment that a rotating medium induces a rotation of the polarisation of the transmitted light.  Player (1976) confirmed that this observation could be accounted for through an application of Maxwells equations in a moving medium. More recently Padgett \textit{et al.} (2006) reasoned that the rotation of the medium turns a transmitted image by the same angle as the polarisation. This is in contrast to the Faraday effect (Faraday 1846), where a static magnetic field in a dielectric medium, parallel to the propagation of light, causes a rotation of the polarisation but not a rotation of a transmitted image. Rotation of the plane of polarisation and image rotation in a rotating medium may be attributed respectively to the spin and orbital angular momentum of light (Allen \textit{et al.} 1999, 2003).  

The first theoretical treatment of this problem was published by Fermi (1923), who considered plane waves and a non-dispersive medium.
The theoretical ana\-lysis of Player (1976) was also restricted to the propagation of plane waves, but took the dispersion of the medium into account. Player assumed that the dielectric response does not depend on the motion of the medium. In our treatment we follow his assumption although a more careful analysis by Nienhuis \textit{et al.} (1992) showed that there will be an effect of the motion on the refractive index for a dispersive medium near to an absorption resonance (see also Baranova \& Zel'dovich (1979) for a discussion on the effect of the Coriolis force on the refractive index). In contrast to Player we allow for more general electromagnetic fields that can carry orbital angular momentum (OAM). This leads to an additional term in our wave equation, which corresponds to a Fresnel drag term familiar from analysis of uniform motion. For a rotating medium, however, this drag leads to a rotational shift of the image. The propagation of light in a rotating medium thus involves both spin angular momentum (SAM) and OAM.
We solve the wave equation for circularly polarised Bessel beams and consider two different superpositions of such Bessel beams to quantify the effects of both polarisation and image rotation. For rotation of the polarisation we examine a superposition of left- and right-circularly polarised Bessel beams carrying the same amount of OAM. For image rotation we consider a superposition of Bessel beams with the same circular polarisation but opposite OAM values. Such a superposition creates an intensity pattern with lobes or `petals'. In both cases the constituent Bessel beams propagate differently in the medium, which leads to a change in their relative phase. This is the origin of the rotation of both the polarisation and the transmitted image. For both phenomena we derive an expression for the angle per unit length of dielectric through which the image or the polarisation is rotated.

The significance of the total angular momentum can be most easily seen in the wave equation for the propagation of light in a rotating medium. We derive this wave equation in section \ref{sec:waveeq}. In the remaining sections we calculate the rotation of polarisation (section \ref{sec:polrot}) and the image rotation (section \ref{sec:imgrot}) and reveal their common form.

%
% WAVE EQUATIONS
%

\section{Wave equations}
\label{sec:waveeq}

The wave equation for a general electric displacment $\mathbf{D}$ in a rigid dielectric medium rotating with angular velocity $\mathbf{\Omega}$ is given by:
\begin{equation}
\label{eq:finalwaveeq}
-\nabla^2 \mathbf{D} = - \epsilon(\omega')  \ddot{\mathbf{D}} + 2 [\epsilon(\omega') - 1] 
\left[ \mathbf{\Omega} \times \dot{\mathbf{D}} - (\mathbf{v} \cdot \nabla) \dot{\mathbf{D}}  \right]. 
\end{equation}
An analoguous wave equation can be derived for the magnetic induction $\mathbf{B}$.
Compared to the form derived by Player (1976), who considered the special case of a plane wave propagating along the direction of $\mathbf{\Omega}$, these wave equations contain an additional term $2 [\epsilon(\omega') - 1] (\mathbf{v} \cdot \nabla) \dot{\mathbf{D}} $. This term is responsible for the Fresnel drag effect which modifies the speed of light in a moving medium (McCrea 1954; Barton 1999; Rindler 2001). In the following we will derive this wave equation for the electric displacement.

Our analysis starts with the same considerations as Player (1976), by introducing a rest frame and a moving frame. In the rest frame the dielectric medium rotates with an angular velocity $\mathbf{v} = \mathbf{\Omega} \times \mathbf{r}$ and in the moving frame the medium is at rest. We restrict our analysis to small velocities with $v \ll c$ and use Maxwell's equations in both reference frames (Landau \& Lifshitz 1975). For the medium at rest we assume the following constitutive relations:
\begin{subequations}
\label{eq:constitutive}
\begin{align}
\mathbf{D}' & = \epsilon(\omega') \mathbf{E}',\\
\mathbf{B}' & = \mathbf{H}',
\end{align}
\end{subequations}
where we have used primes to denote the fields and their frequency $\omega'$ in the moving frame. 
The fields in the moving frame can be  expressed in the rest frame by a Lorentz transformation (Stratton 1941; Jackson 1998), which gives to first order in $v/c$:
\begin{subequations}
\label{eq:transforms}
\begin{align}
\mathbf{D}' & = \mathbf{D} + \mathbf{v} \times \mathbf{H}, \\
\mathbf{B}' & = \mathbf{B} - \mathbf{v} \times \mathbf{E}, \\
\mathbf{E}' & = \mathbf{E} + \mathbf{v} \times \mathbf{B}, \\
\mathbf{H}' & = \mathbf{H} - \mathbf{v} \times \mathbf{D},
\end{align}
\end{subequations}
where we have set $c=1$ and work with units in which $\epsilon_0 = \mu_0 = 1$. The two constitutive relation in (\ref{eq:constitutive}) in the rest frame are thus given by
\begin{subequations}
\begin{align}
\mathbf{D} + \mathbf{v} \times \mathbf{H} & = \epsilon(\omega') \left( \mathbf{E} + \mathbf{v} \times \mathbf{B} \right), \\
\mathbf{B} - \mathbf{v} \times \mathbf{E} & = \mathbf{H} - \mathbf{v} \times \mathbf{D}.
\end{align}
\end{subequations}
The dielectric constant is still given as a function of the frequency in the moving frame. We also assume that the dielectric constant depends only on the frequency and is otherwise independent of the state of motion of the medium. On combining these two equations we can express $\mathbf{D}$ and $\mathbf{B}$ with the two other fields
$\mathbf{E}$ and $\mathbf{H}$ to the first order in $v$:
\begin{subequations}
\label{eq:dbfirstorder}
\begin{align}
\mathbf{D} & = \epsilon(\omega') \mathbf{E} +  [\epsilon(\omega') - 1] \mathbf{v} \times \mathbf{H}, \label{eq:dfirstorder}\\
\mathbf{B} & = \mathbf{H} - [ \epsilon(\omega') - 1] \mathbf{v} \times \mathbf{E}. \label{eq:bfirstorder}
\end{align}
\end{subequations} 
After taking the curl of (\ref{eq:dfirstorder}) we can use the Maxwell equation $\nabla \times \mathbf{E} = -\dot{\mathbf{B}}$ and express $\dot{\mathbf{B}}$, with the help of (\ref{eq:bfirstorder}), in terms of $\dot{\mathbf{H}}$ and $\dot{\mathbf{E}}$. If we assume $\mathbf{v}$ to be constant (see  \ref{app:acceleration}), as in Player's paper (Player, 1976) this yields 
\begin{equation}
\label{eq:vsteady}
\nabla \times \mathbf{D} = - \epsilon(\omega') \dot{\mathbf{H}} + \epsilon(\omega') [\epsilon(\omega') -1] \mathbf{v} \times \dot{\mathbf{E}} +  [\epsilon(\omega') -1] \nabla \times ( \mathbf{v} \times \mathbf{H}).
\end{equation}
It follows from (\ref{eq:dfirstorder}) that $\epsilon(\omega') \mathbf{v} \times \mathbf{E} = \mathbf{v} \times \mathbf{D}$, to the first order in $\mathbf{v}$, and so we can rewrite (\ref{eq:vsteady}) as:
\begin{equation}
\label{eq:curld}
\nabla \times \mathbf{D} = - \epsilon(\omega') \dot{\mathbf{H}} + [\epsilon(\omega') -1] \mathbf{v} \times \dot{\mathbf{D}} +  [\epsilon(\omega') -1] \nabla \times ( \mathbf{v} \times \mathbf{H}).
\end{equation}
We can now take the curl of (\ref{eq:curld}) to obtain a wave equation for $\mathbf{D}$, as $\nabla \times \nabla \times \mathbf{D} =  - \nabla^2 \mathbf{D}$ for $\nabla \cdot \mathbf{D} = 0$, and the curl of $\dot{\mathbf{H}}$ is given by $\nabla \times \dot{\mathbf{H}} = \ddot{\mathbf{D}}$. In order to express the curl of the vector products  we use the identity $\nabla \times (\mathbf{a} \times \mathbf{b}) = \partial_i b_i \mathbf{a} -  \partial_i a_i \mathbf{b}$, where the doubly occurring index denotes a summation over the Cartesian components. The operator $\partial_i$ represents differentiation with respect to the $i$th component and acts on the whole product which gives rise to terms containing the divergences of $\mathbf{v}, \mathbf{D}$ and $\mathbf{H}$. These terms are either zero, because $\nabla \cdot \mathbf{v} = 0$ and $\nabla \cdot \mathbf{D} = 0$ or they lead to terms which are of second order in $\mathbf{v}$ and therefore negligible. The wave equation for $\mathbf{D}$ is thus given by
\begin{equation}
\label{eq:waveeq}
\begin{split}
-\nabla^2 \mathbf{D} & = - \epsilon(\omega')  \ddot{\mathbf{D}} + [\epsilon(\omega') - 1] 
\left[ (\dot{\mathbf{D}} \cdot \nabla) \mathbf{v} - (\mathbf{v} \cdot \nabla) \dot{\mathbf{D}}  \right]  \\
& + [\epsilon(\omega') - 1] \nabla \times \left[ (\mathbf{H} \cdot \nabla) \mathbf{v} - (\mathbf{v} \cdot \nabla) \mathbf{H} \right].
\end{split}
\end{equation}
For a rotation $\mathbf{v} = \mathbf{\Omega} \times \mathbf{r}$ we can specify terms of the form $(\mathbf{a} \cdot \nabla) \mathbf{v}$ by expressing the components of the velocity $\mathbf{v}$  using the Levi-Civitta symbol $\varepsilon_{ijk}$ as  $v_i = \varepsilon_{ijk} \Omega_j r_k$. The components of $
(\mathbf{a} \cdot \nabla) \mathbf{v}$ are thus given by
\begin{equation}
\label{eq:directderiv}
 \left[ (\mathbf{a} \cdot \nabla) \mathbf{v} \right]_i = a_l \partial_l \varepsilon_{ijk} \Omega_j r_k = 
 a_l \varepsilon_{ijk} \Omega_j \delta_{lk} = \left[ \mathbf{\Omega} \times \mathbf{a} \right]_i.
\end{equation}\
If we use the results from (\ref{eq:directderiv}) in (\ref{eq:waveeq}) we find for $\nabla^2 \mathbf{D}$:
\begin{equation}
\begin{split}
\label{eq:waveeqrot}
-\nabla^2 \mathbf{D} & = - \epsilon(\omega')  \ddot{\mathbf{D}} + [\epsilon(\omega') - 1] 
\left[ \mathbf{\Omega} \times \dot{\mathbf{D}} - (\mathbf{v} \cdot \nabla) \dot{\mathbf{D}}  \right] \\ 
& + [\epsilon(\omega') - 1] \nabla \times \left[ \mathbf{\Omega} \times \mathbf{H} - (\mathbf{v} \cdot \nabla) \mathbf{H} \right].
\end{split}
\end{equation}
The curl of the last bracket requires some some additional calculations. The first term is given by:
\begin{equation}
\nabla \times \left( \mathbf{\Omega} \times \mathbf{H} \right) =  \left( \nabla \cdot \mathbf{H} \right) \mathbf{\Omega} - \left( \mathbf{\Omega} \cdot \nabla \right) \mathbf{H},
\end{equation}
and the second term can be written as:
\begin{equation}
\nabla \times \left( \mathbf{v} \cdot \nabla \right) \mathbf{H} = \mathbf{\Omega} \left(  \nabla \cdot \mathbf{H} \right) - \nabla \left( \mathbf{\Omega} \cdot \mathbf{H} \right) + \left( \mathbf{v} \cdot \nabla \right) \dot{\mathbf{D}},
\end{equation}
where the last term originates from $\nabla \times \mathbf{H}$. The terms containing the divergence of $\mathbf{H}$ cancel and the term $\left( \mathbf{v} \cdot \nabla \right) \dot{\mathbf{D}}$ can be added to the second term in (\ref{eq:waveeqrot}). The two remaining terms $-\left( \mathbf{\Omega} \cdot \nabla \right) \mathbf{H}$ and $\nabla \left( \mathbf{\Omega} \cdot \mathbf{H} \right)$ together give
$\mathbf{\Omega} \times \dot{\mathbf{D}}$:
\begin{equation}
\mathbf{\Omega} \times \dot{\mathbf{D}} = \mathbf{\Omega} \times \left( \nabla \times \mathbf{H} \right) = 
\nabla \left( \mathbf{\Omega} \cdot \mathbf{H} \right) - \left( \mathbf{\Omega} \cdot \nabla \right) \mathbf{H}.
\end{equation}
This concludes the derivation of the wave equation (\ref{eq:finalwaveeq}). It is possible to derive the same wave equation for $\mathbf{B}$ using similar methods.
 
For a rotation around the $z$ axis with constant angular velocity $\mathbf{\Omega} = \Omega \mathbf{e}_z$, the directional derivative $\mathbf{v} \cdot \nabla$ is proportional to an azimuthal derivative, as $\mathbf{v} \cdot \nabla = \mathbf{\Omega} \times \mathbf{r} \cdot \nabla = \Omega \partial_\phi$. This allows us to identify the two terms $\mathbf{\Omega} \times \dot{\mathbf{D}}$ and $\Omega \partial_\phi \dot{\mathbf{D}}$ in the wave equation
\begin{equation}
\label{eq:phiwaveeq}
-\nabla^2 \mathbf{D} = - \epsilon(\omega')  \ddot{\mathbf{D}} + 2 [\epsilon(\omega') - 1] 
\left[ \mathbf{\Omega} \times \dot{\mathbf{D}} - \Omega \partial_\phi \dot{\mathbf{D}}  \right]
\end{equation}
as the polarisation rotation and rotary Fresnel drag terms, respectively. Player's derivation does not contain the term proportional to $\partial_\phi \dot{\mathbf{D}}$ because he treated only the case of a plane wave propagating in the $z$-direction and for such fields $\mathbf{D}$ is independent of $\phi$.

On substituting a monochromatic ansatz of the form $\mathbf{D} = \mathbf{D}_0 \exp(-\rmi\omega t)$ into 
(\ref{eq:phiwaveeq}), where $\omega$ is the optical angular frequency in the rest frame, we obtain:
\begin{equation}
\label{eq:monowaveeq}
-\nabla^2 \mathbf{D}_0 = \epsilon(\omega') \omega^2 \mathbf{D}_0 - 2 [ \epsilon(\omega') - 1 ] \omega \Omega \left[ \rmi \mathbf{e}_z \times \mathbf{D}_0 - \rmi \partial_\phi \mathbf{D}_0 \right].
\end{equation}
If we make an ansatz for $\mathbf{D}_0$ with a general polarisation given by the complex numbers $\alpha$ and $\beta$ (with $|\alpha|^2 + |\beta|^2 = 1$) in the form of $\mathbf{D}_0 = (\alpha \mathbf{e}_x + \beta \mathbf{e}_y) \mathcal{D} + \mathcal{D}_z \mathbf{e}_z$, we find that the $x$ and $y$ components of the wave equation (\ref{eq:monowaveeq}) decouple if $\beta = \pm \rmi \alpha$ corresponding to left- and right-circularly polarised light respectively. If we restrict the solutions to these two cases we can write the wave equation as:
\begin{equation}
\nabla^2 \mathcal{D}= - \epsilon(\omega') \omega^2 \mathcal{D}+ 2 [ \epsilon(\omega') -1 ] \omega \Omega \left( \pm 1 - \rmi \partial_\phi \right) \mathcal{D},
\end{equation}
where the plus sign refers to left-circular polarisation and the minus sign to right-circular polarisation. We can then identify $\pm 1$ as the extreme values of the variable $\sigma$ which corresponds to the circular polarisation or SAM of the light beam. Similarly we can identify $-\rmi \partial_\phi = L_z$ as the OAM operator, so that the wave equation contains a term which depends on the total angular momentum $\sigma + L_z$:
\begin{equation}
\label{eq:transversewaveeq}
\nabla^2 \mathcal{D}= - \epsilon(\omega') \omega^2 \mathcal{D}+ 2 [ \epsilon(\omega') -1 ] \omega \Omega \left( \sigma + L_z \right) \mathcal{D}.
\end{equation}
We shall see that it is the dependence on the optical angular momentum that is responsible for the rotation of both the polarisation and of a transmitted image.

%
% SPECIFIC ROTARY POWER
%

\section{Specific rotary power}
\label{sec:polrot}

The rotation of the polarisation arises from the difference in the refractive indices for left- and right-circularly polarised light. The angle per unit length by which the polarisation is rotated is called the specific rotary power. For an optically active medium at rest the specific rotary power is characteristic for a given material, but from (\ref{eq:transversewaveeq}) it can be seen that light propagates differently in a rotating medium, depending on whether the circular polarisation turns in the same rotation sense as the dielectric or in the opposite sense. This phenomenon is described by the effective specific rotary power (Jones 1976; Player 1976). 

The specific rotary power, defined as (Fowles 1975):
\begin{equation}
\label{eq:rotarypow}
\delta_{\mathrm{pol}}(\omega) = \left( n_r(\omega) - n_l(\omega) \right) \frac{\pi}{\lambda} =  \left( n_r(\omega) - n_l(\omega) \right) \frac{1}{2} \frac{\omega}{c},
\end{equation}
is the angle of rotation of the plane of polarisation in an optical active medium. Here, the indices $r$ and $l$ refer to right- and left-circularly polarised light. It was convenient to set $c=1$ for our derivation in section \ref{sec:waveeq} but we reintroduce it here to facilitate the calculation of measurable quantities. 
In order to illustrate the effect of the OAM of light we choose a Bessel beam as an ansatz for the electrical displacement in the $x-y$ plane:
\begin{equation}
\mathcal{D}= J_m(\kappa \rho) \exp(\rmi m \phi) \exp(\rmi k_z z),
\end{equation}
where $\kappa$ and $k_z$ are the transverse and longitudinal components of the wavevector. Bessel beams of this form carry OAM of $m\hbar$ per photon (Allen \textit{et al.} 1992, 1999, 2003). Substituting the Bessel beam ansatz in the wave equation (\ref{eq:transversewaveeq}) yields the following result for the overall wavenumber $k = \sqrt{\kappa^2 + k_z^2}$:
\begin{equation}
k_{l/r}^2(\omega) = \epsilon(\omega') \frac{\omega^2}{c^2} - 2[\epsilon(\omega') - 1] \frac{\Omega \omega}{c^2} ( \sigma + m).
\end{equation}
The indices $l$ and $r$ denoting the circular polarisation correspond respectively to $\sigma = 1$ and $\sigma = -1$. With the help of the relations $\epsilon(\omega') = n^2(\omega')$ and $k(\omega) = n(\omega) \omega / c$ we can turn the equation for the wavenumbers into an equation for the effective refractive indices for left- and right-circularly polarised light:
\begin{equation}
\label{eq:refindex}
n^2_{l/r}(\omega) = n^2(\omega') - 2[n^2(\omega') - 1] \frac{\Omega}{\omega} ( \sigma + m).
\end{equation}
Following Player (1976) we assume that $\Omega \ll \omega$ and we can therefore approximate the square root for the refractive indices $n_{l/r}$ by a small parameter expansion to the first order in $\Omega/\omega$:
\begin{equation}
n_{l/r}(\omega) \simeq n(\omega') - \left[ n(\omega') - \frac{1}{n(\omega')} \right] \frac{\Omega}{\omega}  \left( \sigma + m \right).
\end{equation}
The frequency in the moving frame $\omega'$ is different for left- and right-circularly polarised light (Garetz 1981) and, more generally, the azimuthal or rotational Doppler shift is proportional to the total angular momentum $(\sigma + m)$ (Allen \textit{et al.} 1994; Bialynicki-Birula \& Bialynicka-Birula 1997; Courtial \textit{et al.} 1998; Allen \textit{et al.} 2003). For left-circularly polarised light with $\sigma=1$ the frequency is thus $\omega' = \omega - \Omega(1+m)$, and for right-circularly polarised light with $\sigma=-1$ the frequency changes to $\omega' = \omega - \Omega(-1+m)$. Following Player (1976) we expand the refractive index of the dielectric in a Taylor series to calculate the difference $n_r - n_l$:
\begin{subequations}
\label{eq:refindices}
\begin{align}
n_l(\omega) & \simeq  n(\omega)  - \frac{d n}{d \omega} \Omega (1+m) - \left[ n(\omega) - \frac{1}{n(\omega)} \right] \frac{\Omega}{\omega}  \left(1 + m \right), \\
n_r(\omega) & \simeq  n(\omega) -  \frac{d n}{d \omega} \Omega (-1+m) - \left[ n(\omega) - \frac{1}{n(\omega)} \right] \frac{\Omega}{\omega}  \left(-1 + m \right).
\end{align}
\end{subequations}
Higher order derivatives of $n$ become comparable in magnitude if $n'(\omega) \Omega \simeq n(\omega)$. This will only be case for a strongly dispersive medium, such as atomic or molecular gases, near a resonance. For such gaseous media the dielectric response in a rotating medium has to examined more closely (Nienhuis \textit{et al.} 1992). For solid materials, such as a rotating glass rod, and for optical frequencies this condition is not fulfilled and we can neglect higher order derivatives in the expansion (\ref{eq:refindices}). Within Player's assumption that the refractive index is independent of the motion of the medium we find for the effective specific rotary power:
\begin{equation}
\delta_{\mathrm{pol}}(\omega) = \left( \omega n'(\omega) + n(\omega) - \frac{1}{n(\omega)} \right) \frac{\Omega}{c}.
\end{equation}
On introducing the group refractive index $n_g(\omega) = n(\omega) + \omega n'(\omega)$ and the phase refractive index $n_\varphi(\omega) = n(\omega)$, we can rewrite the rotary power as
\begin{equation}
\label{eq:playerrotpow}
\delta_{\mathrm{pol}}(\omega) = \left( n_g(\omega) - n_\varphi^{-1}(\omega) \right) (\Omega/c),
\end{equation}
which is identical to Player's (1976) expression. In this form the specific rotary power (\ref{eq:playerrotpow}) can be used directly with experimental data in the SI unit system.
In the next section we look at image rotation caused by a difference in the effective refractive indices for different values of $m$.

%
% Image rotation
% 

\section{Image rotation}
\label{sec:imgrot}

\begin{figure}
\includegraphics[width=0.49\textwidth]{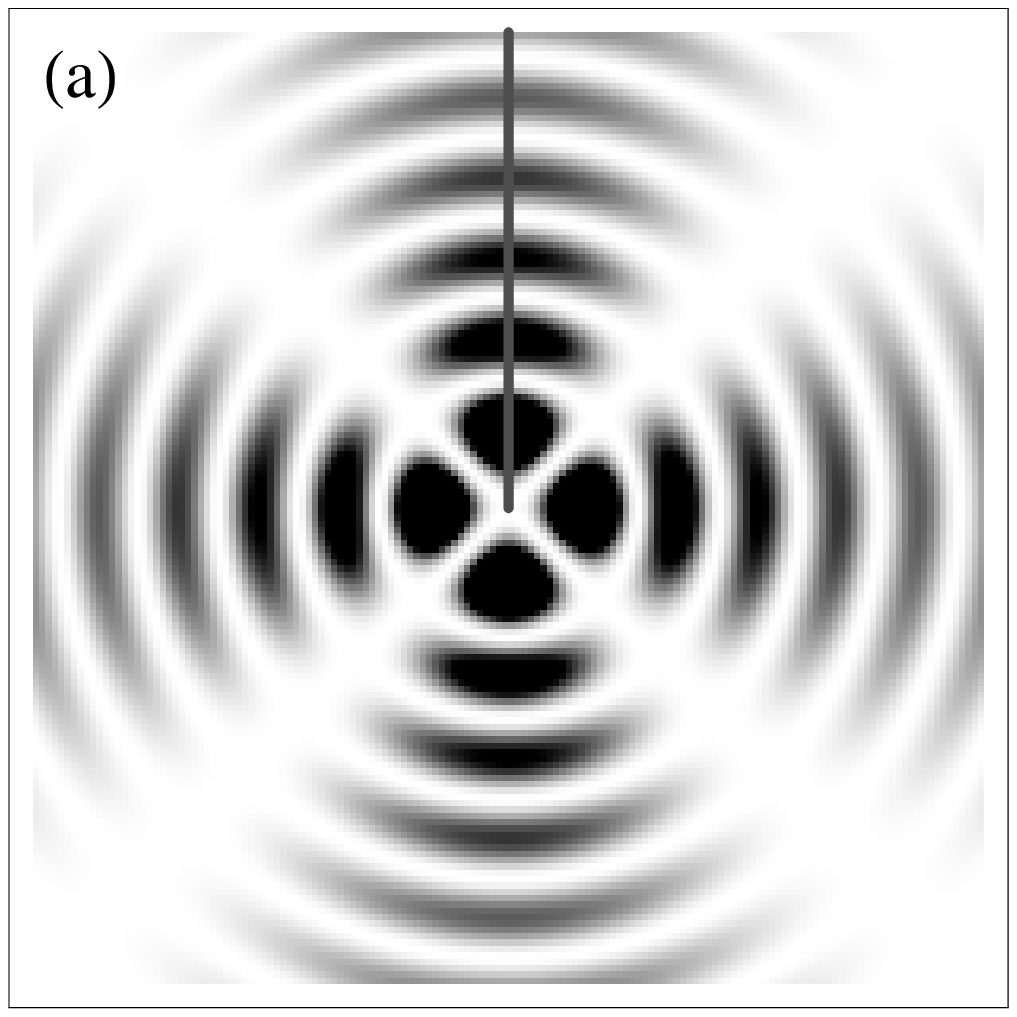}
\includegraphics[width=0.49\textwidth]{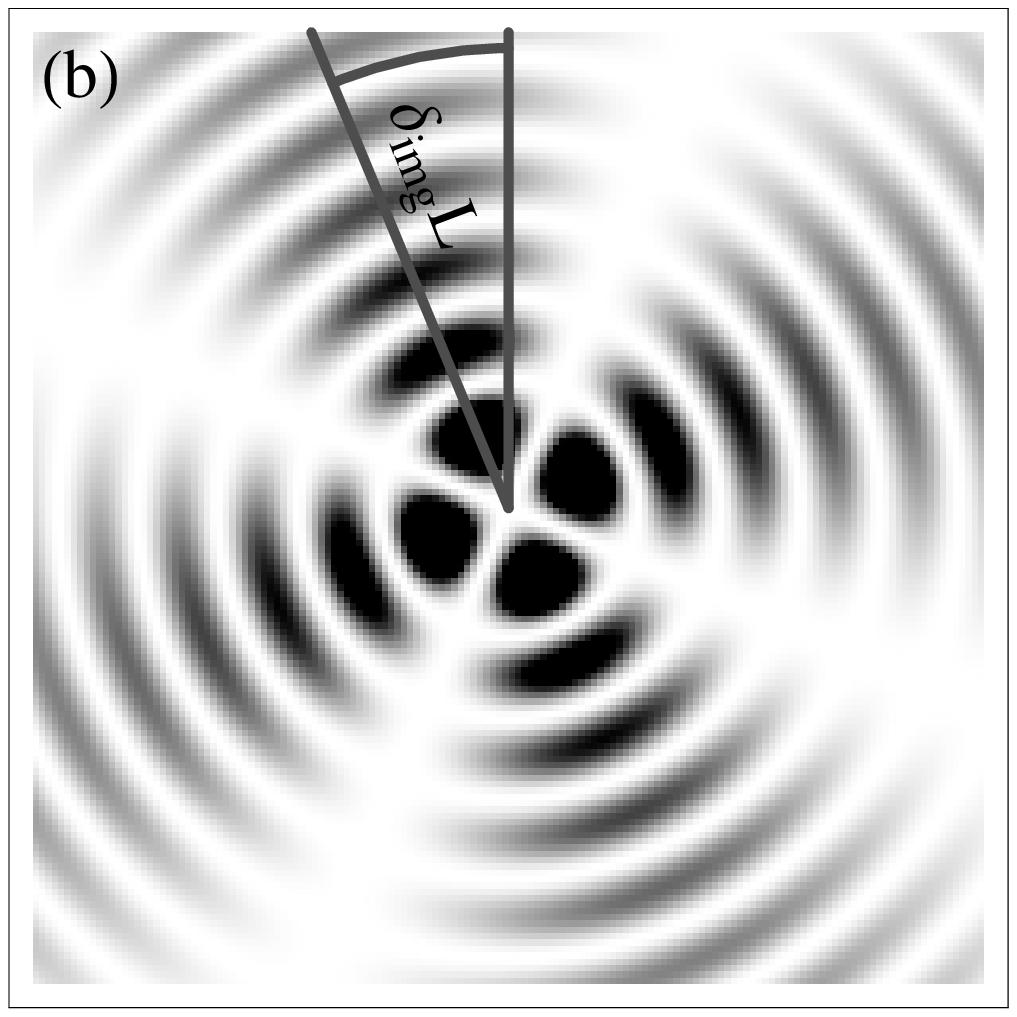}
\caption{\label{fig:petalimage}Image rotation}
\longcaption{Intensity pattern created by the superposition of Bessel beams (a) in (\ref{eq:superposition}) for $m=2$. On propagation the relative phase between the constituent Bessel beams changes which leads to a rotation of the pattern (b). The angle of rotation at a propagation distance $L$ is given by $\delta_{\mathrm{img}} L$.}
\end{figure}

The specific rotary power describes the rotation of the propagation, but we can define, analogously, a rotary power of image rotation. The image can simply be created by the superposition of two light beams carrying different values of OAM which leads to an azimuthal variation of the intensity pattern. In particular we consider an incident superposition of two similarly circularly polarised Bessel beams  with opposite OAM values of the form
\begin{equation}
\label{eq:superposition}
\begin{split}
\mathcal{D}& = \mathcal{D}_+ + \mathcal{D}_- \\
&  = J_m(\kappa \rho) \exp(\rmi m \phi) \exp(\rmi k_z z) + 
J_{-m}(\kappa \rho) \exp(-\rmi m \phi) \exp(\rmi k_z z).
\end{split}  
\end{equation}
Outside the medium the superposition can be written as one Bessel beam with a trigonometric modulation
\begin{equation}
\mathcal{D}= J_m(\kappa \rho) \left( \exp(\rmi m \phi) + (-1)^m \exp(-\rmi m \phi) \right) \exp(\rmi k_z z),
\end{equation}
but inside the medium the effective refractive index is different for the two components of the superposition (Allen \& Padgett 2007). On propagation this leads to phase difference which causes a rotation of the image (see figure \ref{fig:petalimage}). We define 
\begin{equation}
\label{eq:imgrot}
\delta_{\mathrm{img}}(\omega) = \left( n_-(\omega) - n_+(\omega) \right) \frac{1}{2 m} \frac{\omega}{c},
\end{equation}
which is the angle per unit length by which the image is rotated. The factor $m$ in the expression for 
$\delta_{\mathrm{img}}$ appears because of the $\exp(\rmi m \phi)$ and $\exp(-\rmi m \phi)$ phase structure of the interfering beams and the resulting $2m$-fold symmetry of the created image (Pagdett \textit{et al.} 2006).

The different effective refractive indices for the components of the superposition (\ref{eq:superposition})
are given by: 
\begin{equation}
n^2_{+/-}(\omega) = n^2(\omega') - 2[n^2(\omega') - 1] \frac{\Omega}{\omega} ( \sigma \pm m).
\end{equation}
Here, $\sigma$ is fixed in contrast to (\ref{eq:refindex}). The roles of $\sigma$ and $m$ are reversed for the image rotation and the refractive indices for positive and negative OAM  are given by:
\begin{subequations}
\label{eq:rotdiffrefind}
\begin{align}
n_+(\omega) & \simeq  n(\omega)  - \frac{d n}{d \omega} \Omega (\sigma+m) - \left[ n(\omega) - \frac{1}{n(\omega)} \right] \frac{\Omega}{\omega}  \left(\sigma + m \right), \\
n_-(\omega) & \simeq  n(\omega) -  \frac{d n}{d \omega} \Omega (\sigma-m) - \left[ n(\omega) - \frac{1}{n(\omega)} \right] \frac{\Omega}{\omega}  \left(\sigma - m \right).
\end{align}
\end{subequations}
On substituting (\ref{eq:rotdiffrefind}) into (\ref{eq:imgrot}) we find:
\begin{equation}
\delta_{\mathrm{img}} (\omega) = \left( \omega \frac{d n}{d \omega} + n(\omega) - \frac{1}{n(\omega)} \right) \frac{\Omega}{c},
\end{equation}
which can be written in terms of the group and phase refractive indices as:
\begin{equation}
\label{eq:imgrotpow}
\delta_{\mathrm{img}} (\omega) = \left( n_g(\omega) - n_\varphi^{-1}(\omega) \right) (\Omega/c).
\end{equation}
This verifies the reasoning of Padgett \textit{et al.} (2006) that the polarisation and the image are turned by the same amount when passing through a rotating medium. It is the total angular momentum that determines the phase shifts and a linearly polarised image will undergo rotations of both the plane of polarisation and the intensity pattern or image.

%
% CONCLUSION
%

\section{Conclusion}

We have extended a theoretical study by Player (1976) on the propagation of light through a rotating medium to include general electromagnetic fields. In the original analysis Player (1976) showed that the rotation of the polarisation inside a rotating medium can be understood in terms of a difference in the propagation for left- and right-circularly polarised light. Player's (1976) analysis was thus concerned solely with the spin angular momentum (SAM) of light.

Our treatment has shown that the general wave equation has an additional term, which is of the same form as the Fresnel drag term for a uniform motion. In the context of rotating motion, however, this term is connected to the orbital angular momentum (OAM) of the light. By extending the theoretical analysis to include OAM we have been able to attribute polarisation 
rotation and image rotation to SAM and OAM respectively. We have shown that a superposition of Bessel beams with the same OAM but opposite SAM states leads to the rotation of the polarisation, whereas a superposition of Bessel beams with the same SAM and opposite OAM values gives rise to a rotation of the transmitted image.
We have obtained quantitative expressions for the rotation of the polarisation and of the transmitted image  and have verified that both are turned through the same angle, as recently suggested by Padgett \textit{et al.} (2006).

Player (1976) remarked that the derivation by Fermi (1923) appears to be in error. The mistake in Fermi's treatment seems to be in missing the transformation of the magnetic fields. Whereas the change in the electric fields induced by the motion of the medium is explicitly given in terms of the electric polarisation $\mathbf{P}$\footnote{Fermi (1923) denotes the electric polarisation by $\mathbf{S}$}, a similar transformation for the magnetic field is missing. In terms of our derivation this would mean that (\ref{eq:bfirstorder}) changes to $\mathbf{B} = \mathbf{H}$ in the rest frame. This in turn causes that the term $\mathbf{v} \times \dot{\mathbf{D}}$ would be missing in (\ref{eq:vsteady}). This term and 
the term $\nabla \times (\mathbf{v} \times \mathbf{H})$ contribute equally to the wave equation 
(\ref{eq:finalwaveeq}), which explains why Fermi's result for the specific rotary power is smaller than Player's and ours by a factor of two. As pointed out by Player (1976) this missing factor is cancelled by an additional factor of two in Fermi's definition of the specific rotary power.

%
% ACKNOWLEDGEMENTS
%

\begin{acknowledgements}
We would like to thank Amanda Wright and Jonathan Leach whose experiments on this problem motivated our work. This work was supported by the UK Engineering and Physical Sciences Research Council. 
\end{acknowledgements}

%
% APPENDIX
%

\appendix{Accelerated motion}
\label{app:acceleration}

The assumption that $\mathbf{v} = \mathbf{\Omega} \times \mathbf{r}$ is steady is problematic for a rotating motion; if we assume $\Omega$ to be constant over time, then $\dot{\mathbf{v}} = (\mathbf{\Omega} \cdot \mathbf{r} ) \mathbf{\Omega} - \Omega^2 \mathbf{r}$. In principle this would invalidate our initial considerations for the transformation of the electromagnetic fields (\ref{eq:transforms}) which strictly hold only for uniform motion.
Including the time-derivative of $\mathbf{v}$ would lead to additional terms in (\ref{eq:vsteady}) of the form $\epsilon(\omega')[\epsilon(\omega') -1] \dot{\mathbf{v}} \times \mathbf{E}$. If we proceed in taking the curl of this vector product we produce four terms which either can be neglected because they are second order in $v/c$, or they do not contain the time derivative of an optical field. The latter are smaller than terms that do contain a time derivative by $\sim \Omega/\omega$. For our assumption 
$\Omega \ll \omega$ all such terms  are negligible.

\end{document}